\documentclass[10pt,a4paper,twocolumn,showpacs,preprintnumbers,amsmath,amssymb,pra,english,superscriptaddress]{revtex4}

\usepackage[T1]{fontenc}
\usepackage[latin9]{inputenc} 
\usepackage{float}
\usepackage{graphicx}
\usepackage{amssymb}
\usepackage{esint}

\makeatletter
\usepackage{color}

\usepackage{babel}
\makeatother

\begin{document}

\title{Quasicondensation and coherence in the quasi-two-dimensional trapped
Bose gas }

\author{R.~N.~Bisset}
\affiliation{Jack Dodd Centre for Quantum Technology, Department of Physics, University of Otago, New Zealand}
\author{M.~J.~Davis}
\affiliation{ARC Centre of Excellence for Quantum-Atom Optics, School of Physical Sciences,  University of Queensland, Brisbane, QLD 4072, Australia}

\author{T.~P.~Simula}
\affiliation{Department of Physics, Okayama University, Okayama 700-8530, Japan}
\author{P.~B.~Blakie}
\affiliation{Jack Dodd Centre for Quantum Technology, Department of Physics, University of Otago, New Zealand}

\begin{abstract}
We simulate a trapped quasi-two-dimensional Bose gas using a classical
field method. To interpret our results we identify the uniform Berezinskii-Kosterlitz-Thouless (BKT) temperature
$T_{BKT}$ as where the system phase space density satisfies a critical value. We observe that density fluctuations are suppressed in
the system well above $T_{BKT}$ when a quasi-condensate forms as the first occurrence of degeneracy.
At lower temperatures, but still above $T_{BKT}$, we observe the development of appreciable coherence as a prominent finite-size effect, which manifests as bimodality in the momentum distribution of the system.
 At $T_{BKT}$ algebraic decay of off-diagonal correlations occurs near the trap center with an exponent of $0.25$, as expected for the uniform system.
 Our results characterize the low temperature phase diagram for a trapped quasi-2D Bose gas and are consistent with observations made in recent experiments.
\end{abstract}

\maketitle
\section{Introduction}
The physics of two-dimensional systems is very different from what
we observe in the three-dimensional world. 
The Mermin-Wagner-Hohenberg theorem 
\cite{Mermin1966,Hohenberg1967} 
states that thermal fluctuations destroy the long-range order, characteristic of
most phase transitions, in systems of reduced dimension.
However, in systems that support topological
defects, such as vortices, the existence of a quasi-long-range ordered
(quasi-coherent) state was predicted to occur by Berezinskii, Kosterlitz
and Thouless (BKT) \cite{Berezinskii1971,Kosterlitz1973}. 
The BKT superfluid
transition has been experimentally observed in liquid helium thin
films \cite{Bishop1978}, superconducting Josephson-junction arrays
\cite{Resnick1981} and in spin-polarized atomic hydrogen \cite{Safonov}.
More recently evidence for 
the BKT transition  
in a dilute gas of $^{87}$Rb atoms was reported by the ENS group \cite{Stock2005,Hadzibabic2006,Kruger2007}.
However, strong fluctuations, the interplay of harmonic confinement and interactions,
and finite size effects have made predictions for the low temperature
phase diagram of this system the subject of much debate \cite{Popov1983,Petrov2000,Andersen2002,Gies2004,Trombettoni2005,Simula2005}.
Meanfield methods are inapplicable and reliable predictions have only
recently become available from classical field and Quantum Monte Carlo
methods \cite{Simula2006,Holzmann2007b}. 

Limited theoretical understanding has meant that the relationships
between  
bimodality in the
density distribution \cite{Kruger2007}, algebraic decay of phase
coherence \cite{Hadzibabic2006}, and thermal activation of phase
defects \cite{Stock2005,Hadzibabic2006} has been unclear and the
subject of speculation. In particular, a crucial issue currently under debate is whether the
critical point identified using bimodality \cite{Kruger2007} is distinct from
the BKT cross-over found in \cite{Hadzibabic2006}. In contrast to the ENS results a recent experiment by the NIST group \cite{Clade2008} suggests that bimodality occurs at higher temperature (i.e.~at lower phase space density) than the onset of the BKT superfluid.
In this paper
we perform a detailed analysis of the
low temperature properties of a quasi-2D system to explore this issue. Our results support the NIST observations and show that  bimodality occurs at temperatures above the uniform
condition for the BKT transition \cite{Prokofev2001,Prokofev2002}.  
We also show measurable signatures of the degenerate components of the system:  quasi-condensate (suppressed density fluctuations), bimodality (coherence), and superfluidity (algebraic decay of correlations). 
\begin{figure}[H]
\includegraphics[width=3.3in]{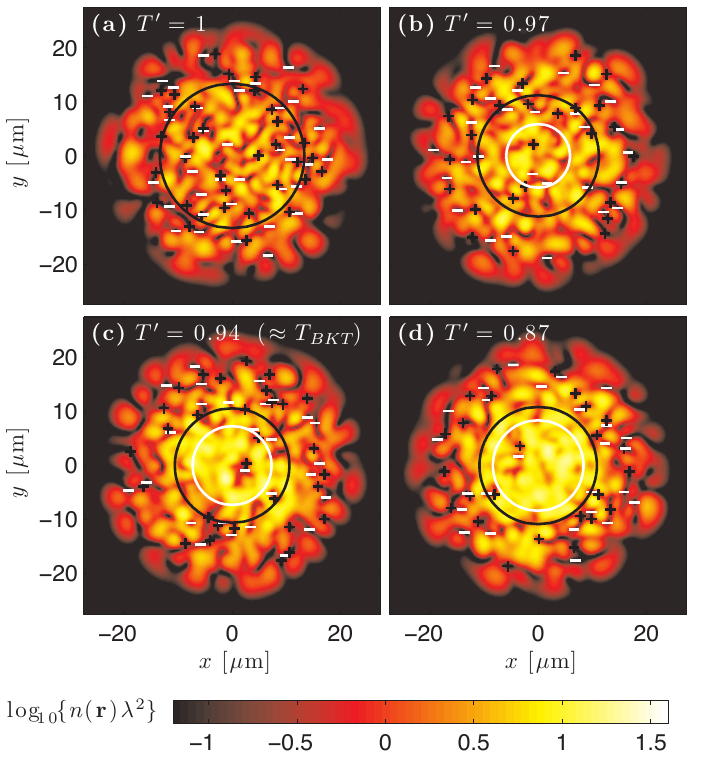}
\caption{\label{fig:vortices} (color online) Instantaneous planar
density of a quasi-2D classical field at a range of temperatures. Inner
white (outer black) circle marks the 1/e-boundary of the condensate
(quasi-condensate) density. Vortices (+) and anti-vortices (-) are
shown. Results indicated for various values of reduced temperature, $T^\prime$ (see Sec.~\ref{RedT}). Other simulation parameters given in the text.}
\end{figure}

To provide insight into the interplay of these degenerate regimes,
Fig.~\ref{fig:vortices} shows four microstates of the quasi-2D trapped
gas 
for a range of temperatures that indicate important features of this
system: In all cases a quasi-condensate (as defined below) is present,
and is the only degenerate component for the highest temperature result
Fig.~\ref{fig:vortices}(a). As the temperature decreases [Fig.~\ref{fig:vortices}(b)] spatial coherence emerges, marked by a finite condensate density. However, in this temperature regime vortices
are prolific and we frequently observe free vortices penetrating the condensate. 
For the lower temperature results
(i.e.~$T\lesssim T_{BKT}$) [see Figs.~\ref{fig:vortices}(c) and
(d)] we see that in the central (condensate) region density fluctuations
are significantly reduced and vortices and anti-vortices are mostly
found paired (i.e.~in close proximity).  

\section{Formalism} 
Here we consider a harmonically trapped
Bose gas described by the Hamiltonian
\begin{equation}
\hat{H}=\int d^{3}x\,\hat{\Psi}^{\dagger}\left\{ H_{sp}+\frac{2\pi a\hbar^{2}}{m}\hat{\Psi}^{\dagger}\hat{\Psi}\right\} \hat{\Psi},\end{equation}
where 
\begin{equation}
H_{sp}=\frac{p^{2}}{2m}+\frac{1}{2}m(\omega_x^2x^2+\omega_y^2y^2+\omega_z^2z^2),
\end{equation}
is the single particle Hamiltonian,
$m$ is the atomic mass, and $a$ is the $s$-wave scattering length.
For $\omega_{z}\gg\omega_{x},\omega_{y}$ the $z$-axis degree of freedom
is frozen out at sufficiently cold temperatures, and the system becomes
quasi-2D. 
The dimensionless 2D coupling constant is $\tilde{g}=\sqrt{8\pi}a/a_{z},$
with $a_{z}=\sqrt{\hbar/m\omega_{z}}$. We will assume that $a_{z}\gg a$
so that the scattering is approximately three-dimensional \cite{Petrov2000},
a condition well-satisfied in the ENS and NIST experiments 
\cite{Stock2005,Hadzibabic2006,Kruger2007,Clade2008}.

In what follows we will take $\mathbf{r}=(x,y)$ to be the 2D position vector, all densities to be areal (i.e.~integrated along $z$), and the trap to be radially
symmetric with $\omega\equiv\omega_{x}=\omega_{y}$ the radial trap
frequency. Our results here are for the case of a quasi-2D system of $^{87}$Rb atoms with trap frequencies $\omega=2\pi\times9.4\,$Hz,
$\omega_{z}=2\pi\times1.88\,$kHz and a coupling constant of $\tilde{g}=0.107$. For reference, in ENS experiments $\tilde{g}\approx0.13$ \cite{Hadzibabic2006}, whereas in the NIST experiments $\tilde{g}\approx$0.02 \cite{Clade2008}.

Our simulation method (see Appendix \ref{AppndPGPE}) is based upon a classical field (denoted by
$\Phi$) representation of the low energy modes of the system. The
remaining high energy modes, that we refer to as the \emph{above region},
are described by a quantum field, $\hat{\psi}$ \cite{Blakie2005a}. Averages of the full field, $\hat{\Psi}=\Phi+\hat{\psi}$,
are obtained by time-averaging the dynamics of the classical field
and ensemble averaging the properties of $\hat{\psi}$ using a Hartree-Fock
meanfield analysis \cite{Simula2006,Davis2006a,Simula2008a}. Because
the classical field approach captures the non-perturbative dynamics
of the low energy modes, it is valid in the critical regime which extends
over a large temperature region in the 2D system and was the basis of the definitive analysis of the uniform 2D Bose gas by Prokof'ev \emph{et al.}, in Ref.~\cite{Prokofev2001}. Here our approach provides a good description of
 the trapped system when: (i)~$\tilde{g}$ is small compared to unity; 
(ii)~all modes of the classical field are appreciably occupied; and
(iii)~the fluctuation region is fully contained within the classical field 
(see Appendix \ref{AppndPGPE}).
Our approach
is computationally efficient and 
suitable for studying non-equilibrium
dynamics, such as the vortex dynamics pictured in Fig.~\ref{fig:vortices}.

We now discuss the various ways of characterizing the 
low temperature components of the quasi-2D system that are useful for interpreting the results of our calculations.

\subsection{Superfluid model of Holzmann \emph{et al.}}
 Recently Holzmann \emph{et al.}~\cite{Holzmann2007b} have
extended a Quantum Monte Carlo method to study this system 
and have proposed a model for the superfluid component
based on a local density application of the uniform results \cite{Prokofev2001,Prokofev2002}.
In detail, they suggested a superfluid density of the form \begin{eqnarray}
n_{sf}(r) & = & \left\{ \begin{array}{cc}
\frac{(m\omega r_{cr})^2}{2\hbar^2\tilde{g}}\left(1-\frac{r^{2}}{r_{cr}^{2}}\right)+\frac{4}{\lambda^2}, &\,\,\,\, r\le r_{cr}\\
0, & \,\,\,\, r>r_{cr}\end{array}\right.\label{eq:nSF1}\end{eqnarray}
where $r_{cr}$ is the radius at which the trapped system density
equals the (temperature dependent) critical value 
\begin{equation}
n_{cr}=\lambda^{-2}\log\left(\frac{C}{\tilde{g}}\right),\label{eq:peakdencond}
\end{equation}
with $C=380\pm3$ and $\lambda=h/\sqrt{2\pi mk_{B}T}$ ($n_{cr}\lambda^{2}$
is the critical phase-space density for the uniform 2D Bose gas to
undergo the BKT transition \cite{Prokofev2001}). To obtain the superfluid density using (\ref{eq:nSF1}) hence requires a comprehensive calculation of the system density (e.g.~using classical field theory and Quantum Monte Carlo) from which $r_{cr}$ can be determined.

In \cite{Holzmann2007b}
Eq. (\ref{eq:nSF1}) was shown to provide a reasonable description
of the moment of inertia obtained directly from Monte Carlo calculations.
However this model has curious behavior at the transition whereby
the superfluid density discontinuously emerges as a sharp spike when
the central (peak) density of the system exceeds $n_{cr}$. 
The temperature where this occurs is denoted $T_{BKT}$ \cite{Holzmann2007b,Holzmann2008a,Bisset2008b}.

\subsection{Quasi-condensate}
Quasi-condensate is the component of the system with
suppressed density fluctuations,
defined by \cite{Prokofev2001}
\begin{equation}
n_{qc}(\mathbf{r})=\sqrt{2\langle\hat{n}(\mathbf{r})\rangle^{2}-
\langle\hat{n}(\mathbf{r})^{2}\rangle},\label{nqc}
\end{equation}
where
$\hat{n}(\mathbf{r})=\hat{\Psi}^{\dagger}(\mathbf{r})\hat{\Psi}(\mathbf{r})$ is
the density operator for the system. Note, that for a normal system with Gaussian density fluctuations $\langle\hat{n}(\mathbf{r})^{2}\rangle=2\langle\hat{n}(\mathbf{r})\rangle^{2}$ so that $n_{qc}=0$.
Definition (\ref{nqc}) was shown to furnish a universal  quantity in the
uniform gas \cite{Prokofev2001}.

\subsection{Coherence} 
There are many ways to define coherence, especially so for an inhomogeneous system, and here we choose to use a measure based on the Penrose-Onsager criterion. Normally this criterion is that a single a macroscopic eigenvalue of the one-body density matrix 
\begin{equation}
G(\mathbf{r},\mathbf{r}')=\langle\hat{\Psi}^{\dagger}(\mathbf{r})\hat{\Psi}(\mathbf{r}')\rangle,
\end{equation}
is equivalent to the condensate number, with respective eigenfunction characterizing the condensate mode. Formally, this analysis is only valid in the thermodynamic limit, and finite size effects make the unique identification of condensate rather fraught -- especially so in the 2D case. Here we choose to use the largest eigenvalue and eigenvector as a measure of coherence, and refer to them as condensate only for clarity \cite{Penrose1956,Blakie2005a}. We denote  the condensate density as $n_{c}(\mathbf{r})$, which provides a spatial characterization of the system coherence (e.g. the condensate mode boundary shown in Fig. \ref{fig:vortices}). 
\begin{figure}
\includegraphics[width=3.2in]{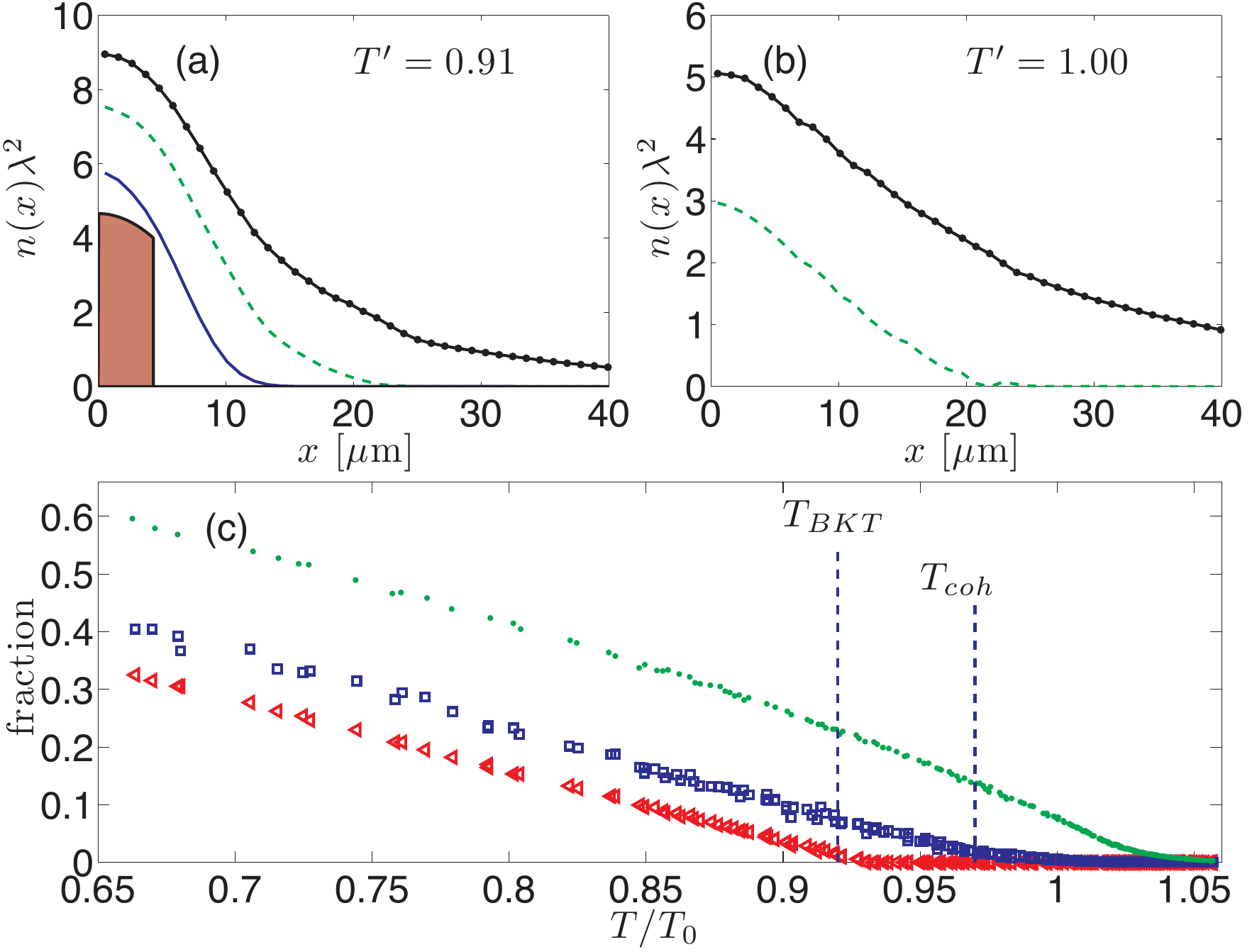}
\caption{\label{fig:phases} (color online) Component densities along the $x$-axis at (a) $T'=0.91$
and (b) $T'=1.0$ : Total density $n(x)$ (dotted-line), quasi-condensate $n_{qc}(x)$ (dashed), condensate
$n_{c}(x)$ (solid) and superfluid $n_{sf}(x)$ (shaded region). (c) Fraction
of atoms in quasi-condensate (dots), condensate (squares) and superfluid
(triangles). 
System (a):  $T=30.6\,$nK and $N=9.9\times10^3$, system (b):  $T=48.5\,$nK and $N=25.9\times10^3$. Other parameters as
discussed in the text.}
\end{figure}
\subsection{Reduced temperature}\label{RedT}
The classical field approach we employ uses a fixed classical region with its total energy as the macroscopic control parameter. As we vary this parameter both the temperature and number of particles vary (see Appendix \ref{AppndPGPE}). To remove leading order effects of varying $N$ we scale all temperatures according to $T'=T/T_0$, where $T_0$ is the quasi-2D ideal gas BEC temperature. We numerically calculate $T_0$ by determining the saturated thermal cloud occupation according to 
\begin{equation}
N(T)=\sum_{j\ne0}\left(e^{(\epsilon_j-\epsilon_0)/k_BT}-1\right)^{-1},
\end{equation}
where $\{\epsilon_j\}$ are the 3D harmonic oscillator energies for our trap. Inverting this relation numerically we obtain $T_0(N)$ - i.e.~the temperature at which the thermal cloud of $N$ atoms is saturated.  The inclusion of quasi-2D effects, i.e. excited $z$-states, is typically important, producing an appreciably lower critical temperature than the pure 2D prediction $T_{2D}=\sqrt{6N\omega_x\omega_z}\hbar/\pi k_B$ \cite{Bagnato1991}.

\section{Results}  
\subsection{Components of the quasi-2D trapped gas}
Time-averaged density profiles are shown in Figs.~\ref{fig:phases}(a) and \ref{fig:phases}(b)
for a system with $\tilde{g}=0.107$.
Figure~\ref{fig:phases}(a) shows the system density when both
quasi-condensate and condensate components are present, whereas 
Fig.~\ref{fig:phases}(b) shows the density at a higher temperature where only a
quasi-condensate component exists.
Figure~\ref{fig:phases}(c) plots the degenerate components
of the system
over a large temperature range. 
At the highest temperatures
we observe a small quasi-condensate fraction, which increases with
decreasing temperature. 
Below $T'\approx0.98$ 
the condensate fraction becomes appreciable (indicated as the \emph{coherence}  temperature  $T_{coh}$ in Fig.~\ref{fig:phases}(c)). In this regime we
commonly see \emph{free} vortices in the condensate 
[e.g.~see Fig.~\ref{fig:vortices}(b)] 
which  have
a profound effect on
its coherence properties. Finally, at $T_{BKT}$ ($T'\approx0.93$) the peak density
at the trap center satisfies Eq.~(\ref{eq:peakdencond}) and the
model, Eq.~(\ref{eq:nSF1}), predicts a non-zero superfluid fraction.
Here we observe the occurrence of vortices in the condensate to be
much less frequent than at higher temperatures, and those found are
usually paired [e.g.~see Fig.~\ref{fig:vortices}(c--d)]. As
the temperature decreases further, the relative difference between 
quasi-condensate, condensate
and superfluid fractions decreases, though they are always clearly
distinguishable.  

An important result of this paper is that the emergence of coherence occurs at higher temperatures than that where the conditions for uniform system superfluidity
are satisfied, $T_{BKT}$, i.e.~the emergence of spatial coherence in
the trapped system occurs before the peak phase space density satisfies
Eq.~(\ref{eq:peakdencond}). We have carried out simulations for $\tilde{g}$
values in the range of $0.02$--$0.107$ and observed similar results. 
\begin{figure}
\includegraphics[width=3.3in]{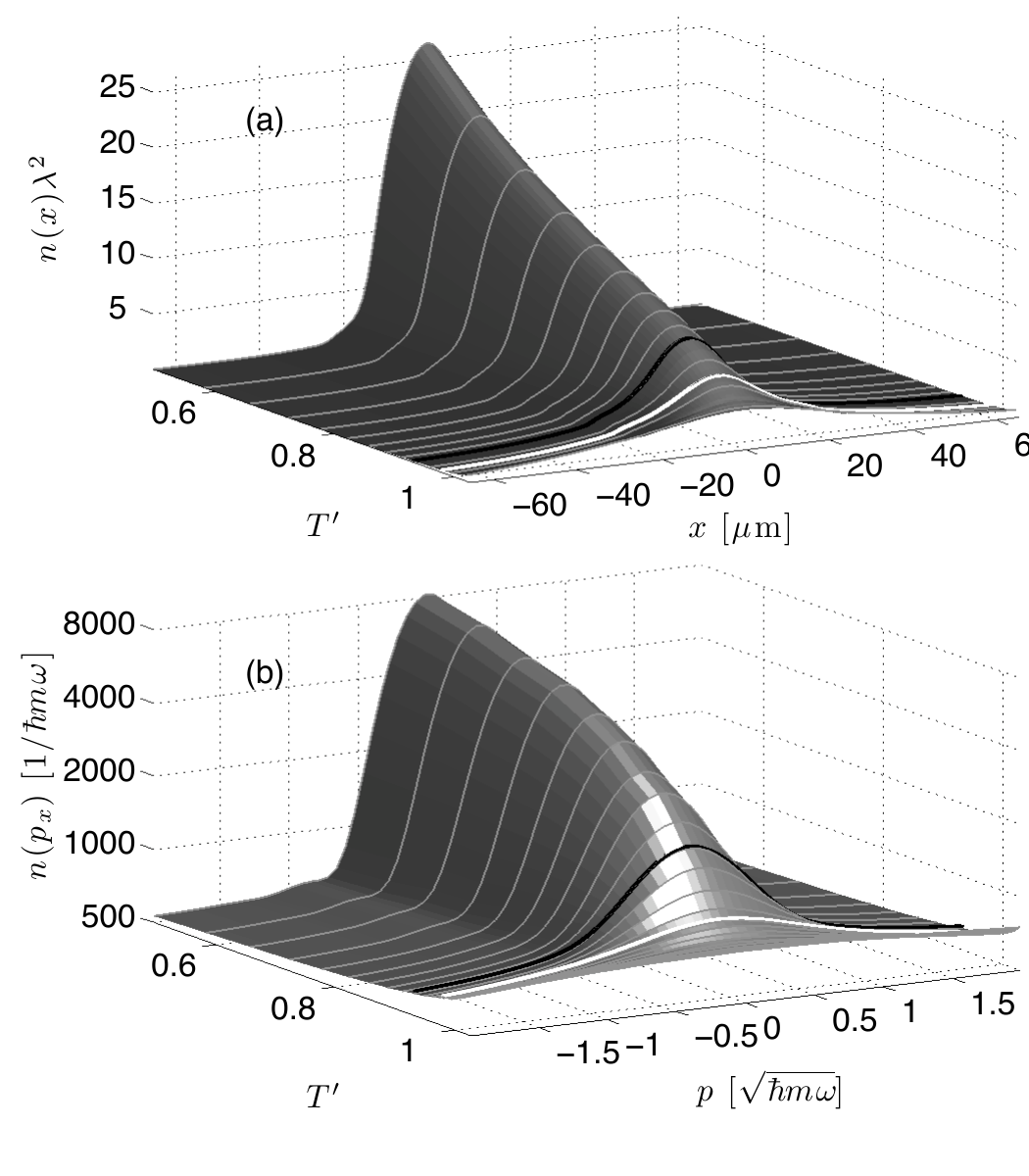}
\caption{\label{fig:qcond} Density distributions versus temperature. (a) Position
phase-space density and (b) momentum space density distributions.
The black (white) curves mark the densities at $T_{BKT}$ ($T_{coh}$)
[also see Fig.~\ref{fig:phases}]. Parameters as in Fig.~\ref{fig:phases}}
\end{figure}

In Fig.~\ref{fig:qcond} we show the
position and momentum density
distributions for our system as a function of temperature.
In Fig.~\ref{fig:qcond}(a)
the position dependent phase-space density is seen to smoothly peak
as the temperature decreases %
\cite{FN1}. We have confirmed that at low temperatures the central density is
well-described by a Thomas-Fermi-type profile of the form $n(r)\sim n(0)-m^2\omega^{2}r^{2}/2\hbar^2\tilde{g}$
(also see \cite{Holzmann2007b}). Additionally, we observe that whenever
appreciable quasi-condensate is present the density profile 
bulges slightly (e.g.~see Fig.~\ref{fig:phases}(b)). Figure~\ref{fig:qcond}(b)
shows that for temperatures below $T'\approx0.98$ (where coherence is observed) that a strong bimodality
is apparent in the momentum space density of the system associated
with the extended spatial coherence arising from the condensate. (Note that the momentum density is shown on a logarithmic scale so that the change from the flat profile at $T'\approx1$ is quite profound). We emphasize 
that this
bimodality is clearly apparent at temperatures well-above $T_{BKT}$. 

\begin{figure}
\includegraphics[width=3.2in]{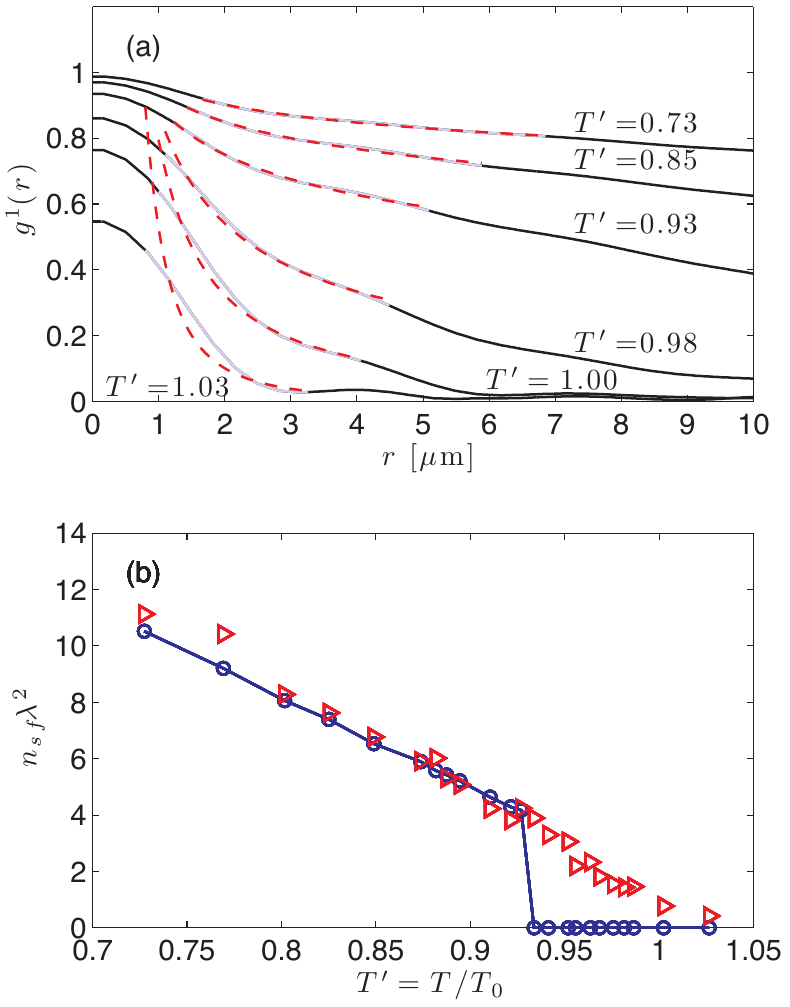}
\caption{\label{fig:G1decay}(color online) (a) $g^{1}(r)$ for various temperatures
(solid lines). Fitting region (light colored segments of curves) and
algebraic fits (dashed) are also shown. (b) Peak superfluid density
determined from fits to algebraic decay (triangles) and from model
density, Eq.~(\ref{eq:nSF1}) (circles). Parameters as in Fig.~\ref{fig:phases} }
\end{figure}

\subsection{Decay of coherence}
 An important prediction of the
BKT theory for the uniform system is that in the superfluid
regime off-diagonal correlations (phase coherence) decay algebraically, 
i.e.~$G(r)\sim r^{-\alpha},$ where $r$ is the spatial separation and the
exponent relates to the superfluid density as $\alpha=1/n_{sf}\lambda^{2}$. At
$T_{BKT}$ the superfluid transition occurs with $n_{sf}\lambda^{2}=4$
(i.e.~$\alpha=0.25$) \cite{Kosterlitz1973} and at temperatures above $T_{BKT}$ the system is normal with exponentially decaying correlations. 

Here we investigate the nature of off-diagonal correlations in the
trapped system \cite{Bezett2008a}. Because of spatial inhomogeneity 
the first order correlation
function depends on centre-of-mass and relative coordinates, and we
therefore choose to examine these correlations symmetrically about
 the trap center.
The normalized correlation function we evaluate is \cite{FN2}
\begin{eqnarray}
g^{1}(\mathbf{r}) & = & 
\frac{\left\langle \Phi^{*}(\mathbf{r}/2)
\Phi(-\mathbf{r}/2)\right\rangle }
{\sqrt{\langle \hat{n}(\mathbf{r}/2)\rangle
\langle\hat{n}(-\mathbf{r}/2)\rangle}},\label{eq:g1}\end{eqnarray}
and typical results 
 for a range of temperatures
are shown in Fig.~\ref{fig:G1decay}(a). We least-squares fit a model
decay curve of the form $g^{1}(r) \propto 1/r^{\alpha}$
 over the spatial
range $1.2\lambda<r<5\lambda$ to determine the exponent $\alpha$. The lower spatial limit excludes
the short range contribution of normal-component atoms, while the
upper limit restricts the effects of spatial inhomogeneity (typical
size of cloud is of order $30\lambda$). The model fits  in
Fig.~\ref{fig:G1decay}(a) are poor in the three highest temperature
cases, suggesting that the algebraic fit is inappropriate. 
Using the fitted values of $\alpha$ we infer the superfluid
density at the trap center according to $n_{sf}(0)=1/\alpha\lambda^{2}$.  
The results, shown in Fig.~\ref{fig:G1decay}(b), compare well with the
peak of the model superfluid density, Eq.~(\ref{eq:nSF1}), for temperatures
below $T_{BKT}$ ($T'\approx0.93$). At higher temperatures (where $n(0)<n_{cr}$) the
two results disagree, however in this temperature range the algebraic
fit is poor and an exponential fit to $g^1(r)$ appears to be more appropriate. 

These results can provide us with an indication that our identification of the superfluid transition is consistent: Using $\alpha=0.25$ as the condition for identifying the BKT transition and comparing to our numerical fits for $\alpha$ would suggest that $T_{BKT}=(0.89-0.94)T_0$, consistent with $T_{BKT} $ as identified according  to condition (\ref{eq:peakdencond}).

\subsection{Relation to experiments}
Our results can be compared against the recent experiments at ENS and NIST.
Kr\"uger \emph{et al.}~\cite{Kruger2007} (ENS) determined the ``exact
critical point'' by observing bimodality in the density distribution
after expansion. Those results suggested that the occurrence of bimodality coincided with the emergence of superfluidity in the system. However, Clad{\'e} \emph{et al.}~\cite{Clade2008} (NIST),  instead saw bimodality occurring prior to the superfluid transition. It is difficult to directly compare each experiment as they differ in several aspects: the ENS experiment typically involved several 2D systems and a large $\tilde{g}$ value, whereas the NIST experiment isolated a single 2D system with a smaller $\tilde{g}$ value. Our results are for a $\tilde{g}$ value intermediate to both experiments, but suggest that bimodality is distinct from superfluidity. However, as the relative temperature separation between these phenomena is quite small they may be difficult to distinguish in practice if temperature resolution is poor.
 We also note that the ENS experimental procedure was to release
the gas for $22\,$ms of time-of-flight before absorption imaging
of the system density. This time scale is on the order of the characteristic
time for expansion in the weak trap direction ($\omega_{x}\approx2\pi\times9.4\,$Hz).
Thus the measured density distribution along $x$ is in an intermediate
regime not well-represented by the \emph{in situ} spatial or momentum
densities, but is in some sense 
a convolution of the features of both. A semiclassical analysis of this expansion is presented in \cite{Hadzibabic2008a}. While the quasi-condensate exhibits a certain amount of spatial bulging, our conjecture is that the sharp bimodal feature seen in the experiment (Fig.~2 of \cite{Kruger2007}) is almost certainly due to the onset
of coherence [see Fig.~\ref{fig:qcond}(b)]. This could
be verified experimentally using longer expansion times to more clearly reveal the system momentum distribution. 

In an earlier experiment \cite{Hadzibabic2006} the BKT transition
was identified by using a heterodyning scheme to determine when off-diagonal correlations in the central
region of the system had an algebraic decay coefficient of $\alpha\approx0.25$.
Our results in Fig.~\ref{fig:G1decay} show
that $\alpha=0.25$  occurs where the peak density satisfies
Eq.~(\ref{eq:peakdencond}), which we have used to define $T_{BKT}$.
The NIST group has developed a more precise procedure for measuring spatial coherence \cite{Clade2008}, which should allow a more systematic investigation of correlations and their relationship to superfluidity and bimodality.

Another issue of concern has been the difference in temperature between the ideal gas condensation temperature, $T_0$, and the emergence of the degeneracy in the interacting system. In Kr\"uger \emph{et al.}~\cite{Kruger2007} the onset of bimodality  was found to occur for atom numbers ``\ldots about 5 times higher than predicted by the semiclassical theory of Bose-Einstein condensation (BEC) in the ideal gas''. This roughly corresponds to the degeneracy temperature being half that of the ideal gas prediction. In a subsequent work the method for determining  temperature was compared with theoretical calculations and was found to underestimate the true temperature by a factor of 0.6-0.7 \cite{Hadzibabic2008a}. Although residual effects of the optical lattice potential used in the ENS experiment make computing  $T_0$ difficult it seems that the temperature at which bimodality is observed is close to $T_0$.  Recent theoretical  work \cite{Holzmann2008a} has also suggested that $T_0$ and $T_{BKT}$ that are quite similar for the regime of the ENS experiment. Clearly additional work examining the critical temperature over a wide parameter regime will be necessary to fully understand this relationship better.

\subsection{Relation to theory of Holzmann and coworkers}
Recently Holzmann and coworkers have performed Quantum Monte-Carlo calculations of the quasi-2D trapped gas \cite{Holzmann2007b}.   Although our results are for a system in a different parameter regime we find qualitatively the same behavior for physical predictions, e.g.~the peak in central density curvature occurring in the transition region (see Fig.~3 of \cite{Holzmann2007b}) and the appearance and behavior of condensate (see Fig.~4 of \cite{Holzmann2007b}). The Quantum Monte Carlo results did not analyze the transition region with the resolution we have in this work, and it is possible that the appearance of coherence before superfluidity is also revealed in their results \cite{FN3}.

Holzmann and coworkers have also developed a rather elegant meanfield theory for predicting superfluid transition  \cite{Holzmann2008a}. While providing a simple means for finding an approximate value for $T_{BKT}$, several aspects of this theory have been scrutinized in \cite{Bisset2008b} and shown to be inconsistent. In particular, the use of (i) the total areal density and (ii) a temperature renormalized interaction strength in Eq.~(\ref{eq:peakdencond}) to identify the superfluid transition instead of the ground transverse mode density and the bare (ground transverse mode) interaction strength, respectively.
Aspects (i) and (ii)  were also used to identify the superfluid transition in the Quantum Monte Carlo results and the aforementioned consistency issues may have contributed additional uncertainty in the transition temperature \cite{Holzmann2007b}.
Because the system we consider here is more two-dimensional than that in \cite{Holzmann2007b}  \cite{FN4}, the distinction between total and ground transverse mode areal densities is negligible for our results. However, in general the ground transverse mode areal density is the relevant quantity to be used in expression (\ref{eq:peakdencond}) (see \cite{Bisset2008b} for additional discussion). Additionally, we note that expression (\ref{eq:peakdencond}) is valid in the thermodynamic limit and it is not clear how large finite-size shifts might be.  Clearly there is a need for more theoretical calculations in the transition region to better clarify the relationship between coherence and superfluidity.

\section{Conclusions}
We have provided a comprehensive analysis of the low-temperature
properties of the trapped 2D Bose gas and have elucidated the roles of
condensation and quasi-condensation.   A
model for the superfluid density, and the temperature $T_{BKT}$, was proposed  in \cite{Holzmann2007b}, and shown to provide a reasonable description
of the system's moment of inertia. Our results show that in the
finite system coherence can emerge at temperatures above $T_{BKT}$. In this regime it is not clear if the system has some finite-size induced superfluidity arising from the condensate, as the central coherent region is often penetrated by vortices  [e.g.~see Fig.~\ref{fig:vortices}(b)] which may act to destroy its superfluidity.  These observations appear to be in agreement with  the results presented in \cite{Clade2008}, and suggest that the strong fluctuations seen experimentally in the initial bimodal state (for $T>T_{BKT}$) are associated with vortices.
Our calculations of the off-diagonal correlations reveal
that extended coherence emerges at temperature well-above $T_{BKT}$, coinciding with the appearance of bimodality in the momentum distribution, whereas the field develops algebraically decaying correlations, accurately described by the uniform  theory of the superfluid system, at lower temperatures $T\lesssim T_{BKT}$.   

As shown in Fig.~\ref{fig:vortices}, the classical field microstates 
describe the creation of vortices and their ensuing dynamics. 
Future work will clarify the nature of vortex pairing, their typical production rates and lifetime,  to better understand their role in the equilibrium state.

\subsection*{Acknowledgments}
The authors acknowledge valuable discussions with the ENS
and NIST groups. PBB and RNB are supported by NZ-FRST contract NERF-UOOX0703.
TPS acknowledges JSPS support. MJD acknowledges the Australian Research Council  support. 

 \appendix
\section{Review of PGPE method and validity conditions}\label{AppndPGPE}
 In this work we use the classical field method known as the projected Gross-Pitaevskii equation (PGPE) formalism, as summarized in Refs.~\cite{Blakie2005a,Davis2006a,Simula2008a,Simula2006,Bezett2008a,cfieldRev2008,Bradley,Bezett2009a}.
In particular, in Refs.~\cite{Simula2006,Simula2008a} this formalism
is developed for application to the quasi-2D trapped Bose gas. We
briefly summarise this method here before discussing details and validity
conditions of our calculations.

\subsection{Use of PGPE for determining equilibrium properties}
A randomized classical field state ($\Phi$) of definite energy, given
by the functional\begin{equation}
E[\Phi]=\int d^{3}x\,\left\{\Phi^{*}H_{sp}\Phi+\frac{2\pi a\hbar^{2}}{m}|\Phi|^{4}\right\},\label{eq:Efunc}\end{equation}
is constructed according to the procedure discussed in Sec. V of Ref.~\cite{Blakie2008e}.
This state is then evolved according to the PGPE\begin{equation}
i\hbar\frac{\partial\Phi}{\partial t}=H_{sp}\Phi+\mathcal{P}\left\{\frac{4\pi a\hbar^{2}}{m}|\Phi|^{2}\Phi\right\},\label{eq:PGPE}\end{equation}
where the projector, $\mathcal{P}$, limits the occupation of the
field to single particle harmonic oscillator modes below a specified
(single particle) energy cutoff $\epsilon_{\rm cut}$. The PGPE (\ref{eq:PGPE})
has been demonstrated to be ergodic so that, after the initial field thermalizes, time averaging
can be used to obtain equilibrium quantities such as density profiles
\cite{Blakie2005a}, correlation functions \cite{Blakie2005a,Bezett2008a,Simula2008a},
and the temperature and chemical potential \cite{Davis2003a,Davis2005a}.

\subsection{Validity conditions}
The classical field method provides an accurate description of the
low energy component of the quasi-2D trapped Bose gas as long as three
conditions can be satisfied (see sections 2 and 3 of \cite{cfieldRev2008} for further details):
\begin{enumerate}
\item[(i)] That $\tilde{g}$ is small compared to unity: This condition ensures that classical
fluctuations dominate quantum fluctuations. This condition is well-satisfied
in our calculations ($\tilde{g}=0.107$) and in current experiments
(for the ENS group $\tilde{g}\approx0.13$ \cite{Hadzibabic2006}
and for the NIST group $\tilde{g}\approx 0.02$ \cite{Clade2008}).
\item[(ii)] Appreciably occupied modes: In order to apply the classical field
description we require that modes of the classical region are appreciably
occupied, i.e. that the least occupied mode of the classical region
has a mean occupation ($\bar{n}_{\min}$) of order 1 or greater. In practice
we analyze this validity requirement in our simulations by calculating
$\bar{n}_{\min}$, taken as the average the occupation of the highest
energy single particle state (i.e. states with single particle energy
$\epsilon\sim\epsilon_{\rm cut}$) \cite{Blakie2007CR}.
\item[(iii)] Good basis: The energy cutoff $\epsilon_{\rm cut}$ has to be sufficiently large that the single
particle eigenstates provide a good basis for describing the interacting
classical region modes. The condition for this requirement is that
$\epsilon_{\rm cut}'\equiv\epsilon_{\rm cut}-\epsilon_{0}\gtrsim{\hbar^{2}}\tilde{g}n/m$,
where $\epsilon_{0}$ is the single particle ground state energy, and $n$ is the average 2D spatial density of the classical field.
This requirement also ensures that the \emph{fluctuation region }(universal
long wavelength modes) is contained in the classical region. This
condition is also discussed in Ref. \cite{Prokofev2001}.
\end{enumerate}

\subsection{Above cutoff atoms}
Having ensured a valid classical region simulation, we can add the
remaining low-density high-energy particles using a Hartree-Fock description,
as discussed in Ref.~\cite{Simula2008a}. This allows us to calculate
the total number of atoms and the total density profile for the full
quasi-2D Bose gas. As long as the third condition above is well-satisfied,
then these high energy atoms are well-described using meanfield theory
\cite{Prokofev2001}.

\begin{figure}
\includegraphics[width=3.2in]{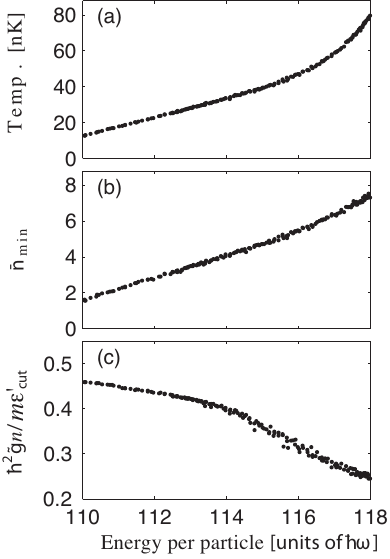}
\caption{
\label{fig:results} Quantities determined from classical field evolution as a
function of the classical field energy per particle.
(a) Simulation temperature. (b) Average occupation of the least occupied
mode. (c) Ratio of interaction energy scale to the energy cutoff.}
\end{figure}

\begin{figure}
\includegraphics[width=3.0in]{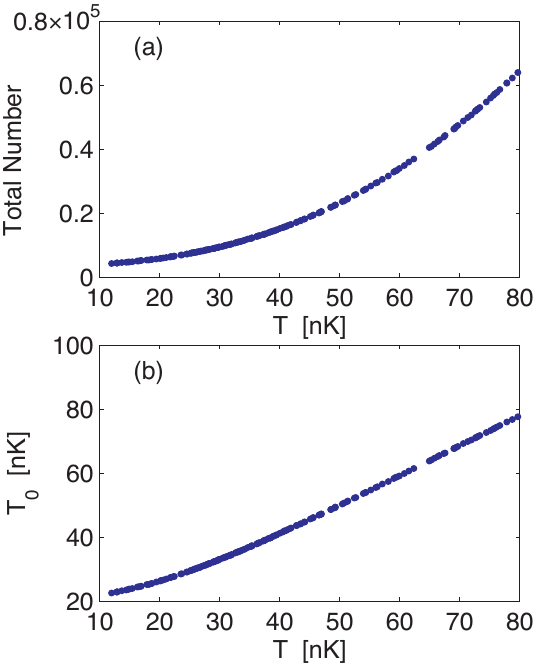}
\caption{
\label{fig:results2} Relevant quantities for our classical field simulations of the quasi-2D trapped Bose gas as a function of temperature.
(a) Total particle number of our results and (b) the respective quasi-2D ideal condensation temperature.}
\end{figure}

\subsection{Parameters for calculation}
In our calculations we have evolved classical field states with a
cut off of $\epsilon_{\rm cut}=125\hbar\omega$, populated with 3750 (classical region) $^{87}$Rb atoms \cite{FN5} for 8.5 seconds
(i.e. over 80 radial trap periods). Allowing the first 40 trap periods
for rethermalization, we ergodically average using 2000 sampled microstates
taken over the last 40 trap periods of evolution.

In Fig.~\ref{fig:results} we show various quantities used to establish
the validity of our calculations. We choose to keep the
classical region size (i.e.~$\epsilon_{\rm cut}$) and number of classical
region atoms fixed, and vary phase space density of the system by
altering the classical region energy as given by Eq.~(\ref{eq:Efunc}).
This change in energy directly affects the temperature, which we determine
using Rugh's dynamical definition of temperature (e.g.~see \cite{Davis2003a,Davis2005a}),
with the results for our parameters shown in Fig.~\ref{fig:results}(a). 

\begin{figure} 
\includegraphics[width=3.0in]{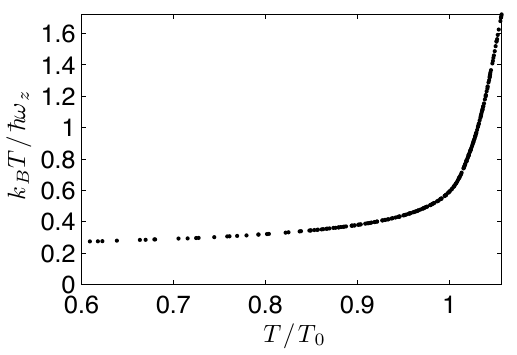}
\caption{Relative thermal activation of $z$ direction of the system as a function of reduced temperature}
\label{fig:results3}
\end{figure}

Of the three validity conditions discussed above, only condition (i)
can be assured \emph{a priori}. Conditions (ii) and (iii) need to be verified
\emph{a posterori }as we now discuss. In Fig.~\ref{fig:results}(b)
the average occupation of the least occupied mode is shown. For all
calculations this is larger than unity (and near the transition it
is typically $\gtrsim3-4$) so that the classical field approach is
well justified for these results. In Fig.~\ref{fig:results}(c) we
show that the ratio $\hbar^{2}\tilde{g}n/m\epsilon_{\rm cut}'$ is small,
ensuring that our cutoff is large enough to provide a good description
of the spectrum and ensuring that the fluctuating region is well contained
within the classical field. In summary Figs.~\ref{fig:results}(b)-(c)
show that our method provides an accurate representation of
the quasi-2D system we have simulated. 
In  Figs.~\ref{fig:results2}(a)-(b) we  show the total  atom number [Fig.~\ref{fig:results2}(a)] and associated ideal condensation temperature, $T_0$, [Fig.~\ref{fig:results2}(b)], which we use to rescale the temperature. Finally in Fig.~\ref{fig:results3} we show the thermal activation of the tight degree of freedom as a function of the reduced temperature.

\begin{thebibliography}{34}
\bibitem{FN1}{The low temperature peaking of $n\lambda^2$ is enhanced relative to $n$ as $\lambda$ increases
with decreasing $T$.}
\bibitem{FN5}{The remaining atoms in energy modes above the classical region are treated using meanfield theory and vary with energy (i.e.~temperature). The total number of atoms in the system are given in Fig.~\ref{fig:results2}(a).}
\end{thebibliography}
\end{document}